\documentclass[12pt]{article}
\usepackage{color}
\usepackage{stmaryrd}
\usepackage{bbm}
\usepackage{mathrsfs}
\usepackage{amsfonts}
\usepackage{amssymb}
\usepackage{amsmath}
\usepackage{amsthm}
\usepackage{cases}
\usepackage{indentfirst}
\usepackage{graphicx}
\usepackage[sort&compress,numbers]{natbib}

\allowdisplaybreaks[4]

\def\be{\begin{eqnarray}} \def\ee{\end{eqnarray}} 
  \def\({\left(} \def\){\right)}
\def\bc{\begin{center}} 
\def\ec{\end{center}}  
\def\bey{\begin{eqnarray*}}\def\eey{\end{eqnarray*}}

\begin{document}

\title{ {\bf Completion of the Ablowitz-Kaup-Newell-Segur
integrable coupling}
\footnotetext{*Corresponding author, Email: yongyangjin@163.com}}
\author{Shoufeng Shen $^{1}$, \quad Chunxia Li $^{2}$, \quad Yongyang Jin $^{1, ~ *}$ \qquad  Wen-Xiu Ma $^{3, ~4,~5}$\\}
\date{}
\maketitle
\begin{center}
\begin{minipage}{135mm}
\noindent {\small $^{1}$ Department of Applied Mathematics, Zhejiang
University of Technology, Hangzhou 310023, PR China}
\\
\noindent {\small $^{2}$ School of Mathematical Sciences, Capital Normal University, Beijing 100048, PR China}\\
\noindent {\small$^3$  College of Mathematics and Systems Science, Shandong University of Science and Technology, Qingdao 266590, PR China}\\
\noindent {\small$^4$
International Institute for Symmetry Analysis and Mathematical Modelling, Department of Mathematical Sciences, North-West University, Mafikeng Campus, Private Bag X2046, Mmabatho 2735, South Africa}
\\
\noindent {\small$^5$  Department of Mathematics and Statistics, University of South Florida, Tampa, Florida 33620-5700, USA}
\end{minipage}
\end{center}

\noindent{\bf Abstract:}

Integrable couplings are associated with non-semisimple Lie algebras. In this paper, we propose a new method to generate new integrable systems through making perturbation in matrix spectral problems for integrable couplings, which is called the `completion process of integrable couplings'.
 As an example, the idea of construction
 is applied to
 the Ablowitz-Kaup-Newell-Segur integrable coupling.
 Each equation in the resulting
hierarchy has a bi-Hamiltonian structure furnished by the component-trace identity.

 \vspace{0.5cm}


 \vspace{0.5cm}


 \vspace{0.5cm}

%-----------------------------------------------------------------------------------------------------
 \numberwithin{equation}{section}
%-----------------------------------------------------------------------------------------------------

\section{Introduction}

Recently, seeking for new integrable couplings  has received considerable attention and formed a pretty important area of research in mathematical physics \cite{mf-csf,ma-maa,gz-jmp, zhang, gzy, ma-jmp,xyc,gz,mxz,mc,gz-pla, gz-csf2,st,gz-csf, mg-mplb, lf,ma-ra,zt,zht,zf,mz,mz-aa,ma-aip1,taz,ma-cam,  ma-aip2, sljy, yyzs}. Integrable
couplings are coupled systems which contain given integrable equations as their sub-systems. Mathematically, for a given integrable equation
$u_{t}=K(u)=K(x, t, u, u_{x}, u_{xx},\cdots)$, its integrable coupling is an enlarged triangular integrable system of the following form
\begin{eqnarray}\label{a1}
\left\{
\begin{array}{l}
u_{t}=K(u),\\
v_{t}=S(u,v).
\end{array}
\right.
\end{eqnarray}
A well-known example of integrable couplings is the first-order perturbation system \cite{mf-csf}
\begin{eqnarray*}
\left\{
\begin{array}{l}
u_{t}=K(u),\\
v_{t}=K'(u)[v],
\end{array}
\right.
\end{eqnarray*}
where $K'(u)[v]$ denotes the Gateaux derivative $K'(u)[v]=\frac{\partial}{\partial \varepsilon}|_{\varepsilon=0}K(u+\varepsilon v, u_x+\varepsilon v_x, \cdots)$. It is known that an arbitrary Lie algebra over a field of characteristic zero has a semi-direct sum structure of a solvable Lie algebra and a semisimple Lie algebra, which is stated by the Levi-Mal'tsev theorem. Therefore, zero curvature equations over semi-direct sums of Lie algebras, i.e., non-semisimple Lie algebras, lay the foundation for generating integrable couplings. Integrable couplings usually show various specific mathematical structures, such as block matrix type Lax representations, bi-Hamiltonian structures, infinitely many symmetries and conservation laws of triangular form.
A general structure of integrable couplings connected with these kinds of algebras has recognized recently and some examples have been presented such as the Ablowitz-Kaup-Newell-Segur (AKNS), Wadati-Konono-Ichikawa (WKI), Kaup-Newll (KN), Korteweg-de Vries, Boiti-Pempinelli-Tu and Volterra integrable couplings \cite{ma-maa,gz-jmp, zhang, gzy, ma-jmp,xyc,gz,mxz,mc,gz-pla, gz-csf2,st,gz-csf, mg-mplb, lf,ma-ra,zt,zht,zf,mz,mz-aa,ma-aip1,taz,ma-cam,  ma-aip2, sljy, yyzs}.

The simplest non-semisimple Lie algebra $\bar {\mathfrak g}$  consists of square matrices of the following block form
\begin{equation*}
M(A_{1},A_{2})=
\left [
\begin{array}{cccc}
  A_{1} & A_{2}\\
  0  &  A_{1}
\end{array}
\right].
\end{equation*}
 $A_{1}$ and $A_{2}$ are two arbitrary square matrices of the same order.  This algebra has two subalgebras $ \tilde{\mathfrak {g}}=\{M(A_1, 0)\}$ and $ \tilde{\mathfrak {g}}_c=\{M(0,A_2)\}$
which form a semi-direct sum: $\bar {\mathfrak {g}}=\tilde{\mathfrak {g}}\inplus \tilde{\mathfrak  {g}}_c$.
The notion of semi-direct sums means that the two subalgebras $\tilde{\mathfrak {g}}$ and  $\tilde{\mathfrak {g}}_c$ satisfy $[\tilde{\mathfrak {g}},\tilde{\mathfrak {g}}_c]\subseteq \tilde{\mathfrak {g}}_c$. We also require the closure property between $\tilde{\mathfrak {g}}$ and  $\tilde{\mathfrak {g}}_c$ under the matrix multiplication: $\tilde{\mathfrak {g}} \tilde{\mathfrak {g}}_c, ~ \tilde{\mathfrak {g}}_c \tilde{\mathfrak {g}} \subseteq \tilde{\mathfrak {g}}_c $.
In what follows, we give a brief account of the procedure for building AKNS integrable coupling associated with $\bar {\mathfrak g}$.

\noindent{\it Step 1:}  One needs to select an appropriate spectral matrix $\bar{U}\equiv \bar{U}(\bar{u},\lambda)$ with the spectral parameter $\lambda$ to form  a spatial spectral problem
\begin{eqnarray}\label{a2}
& & \phi_x=\bar{U}\phi,\qquad \bar{u}=\left[\begin{array} {c}
p \cr
q\cr
r\cr
s
\end{array}\right], \qquad \phi=\left[\begin{array} {c}
\phi_1 \cr
\phi_2\cr
\phi_3\cr
\phi_4
\end{array}\right],
\end{eqnarray}
where
\begin{eqnarray}\label{a3}
& & \bar{U}=\left[\begin{array} {cc}
U &  U_1 \cr
0 & U
\end{array}\right]=\left[\begin{array} {ccccc}
\lambda& p &\vline & 0 & r\cr
q& -\lambda &\vline & s & 0\cr
\hline
0&0 &\vline & \lambda & p\cr
0&0&\vline &q & -\lambda
\end{array}\right].
\end{eqnarray}
In fact, $\phi_x=U\phi$ is nothing but the classical $2\times 2$ AKNS  spatial spectral problem \cite{akns, tu-jmp, tu-jpa, tah}.

\noindent{\it Step 2:}  We construct a particular solution $\bar{W}=\left[\begin{array} {cc} 
W &  W_1 \cr
0 & W
\end{array}\right]$ expressed in terms of Laurent series  to the stationary zero curvature equation  $\bar{W}_x=[\bar{U}, \bar{W}]$, which is used to obtain recursion relations.
One  also needs to prove the localness property for $\bar{W}$ based on the relations.

\noindent{\it Step 3:}  By means of the solution $\bar{W}$  obtained in previous step, we introduce temporal spectral problems $\phi_{t_m}=\bar{V}^{[m]}\phi, ~\bar{V}^{[m]}=(\lambda^{m} \bar{W})_{+}+\bar{\Delta}_m$
 so that the zero curvature equations $ \bar{U}_{t_m}-\bar{V}_x^{[m]}+\left[\bar{U}, \bar{V}^{[m]}\right]=0$
 generate  the AKNS integrable coupling $\bar{u}_{t_m}=\bar{K}_m$.

\noindent{\it Step 4:} Finally, by using the component-trace identity (or the variational identity) \cite{mz-aa}
\begin{eqnarray}\label{a4}
& & \frac{\delta}{\delta u}\int{\rm tr}\left(W\frac{\partial U_1}{\partial \lambda}+W_1\frac{\partial U}{\partial \lambda}\right){\rm d}x=\lambda^{-\gamma}\frac{\partial}{\partial \lambda}\lambda^{\gamma}{\rm tr}\left(W\frac{\partial U_1}{\partial \bar{u}}+W_1\frac{\partial U}{\partial \bar{u}}\right),
\end{eqnarray}
we can furnish bi-Hamiltonian structure
\begin{eqnarray*}
& & \bar{u}_{t_m}=\bar{K}_m=\bar{J}\frac{\delta \bar{\mathcal{H}}_{m}}{\delta u}=\bar{M}\frac{\delta \bar{\mathcal{H}}_{m-1}}{\delta \bar{u}},\qquad m\geq 1,
\end{eqnarray*}
for the obtained AKNS integrable coupling.

In this paper,
we would like to generalize the spatial  spectral problem of AKNS integrable coupling \eqref{a3} by using perturbation technique, namely, adding a nonlinear perturbation term $h$,
\begin{eqnarray}\label{a5}
& & \bar{U}=\left[\begin{array} {cc}
U &  U_1 \cr
0 & U
\end{array}\right]=\left[\begin{array} {ccccc}
\lambda+h& p &\vline & 0 & r\cr
q& -\lambda-h &\vline & s & 0\cr
\hline
0&0 &\vline & \lambda+h & p\cr
0&0&\vline &q & -\lambda-h
\end{array}\right], \qquad h=\epsilon(ps+qr).
\end{eqnarray}
Obviously, this generalized spatial spectral problem is reduced to the case of AKNS integrable coupling \eqref{a3} for $\epsilon=0$.
With the additional nonlinear term $h$, the generalized matrix spectral problem
generates a generalization of the AKNS integrable coupling, which
 takes the form  $
\left\{
\begin{array}{l}
u_{t}=\tilde{K}(u,v),\\
v_{t}=\tilde{S}(u,v).
\end{array}
\right.
$
When $\epsilon=0$,
the resulting integrable system becomes the standard AKNS integrable coupling. In this sense, we call the generalization of integrable couplings the `completion process of integrable couplings'.

The rest of this paper is organized as follows. In Section 2, we will construct a generalization of the AKNS integrable coupling from zero curvature equations, based on the above-mentioned generalized spatial  spectral problem \eqref{a5}.
In Section 3,
Bi-Hamiltonian structure will be furnished by using the
component-trace identity \eqref{a4}, thereby, all the resulting equations in the new hierarchy possess infinitely many
commuting symmetries and conservation laws. For the sake of convenience, we will use the mathematical software Maple to deal with some complicated symbolic computations. The last section is devoted to conclusions and discussions.

\section{Completion of the AKNS integrable coupling}

Now, let us assume that $\bar{W}$ has the following form
\begin{eqnarray}\label{b1}
& & \bar{W}=\left[\begin{array} {cc}
W &  W_1 \cr
0 & W
\end{array}\right]=\left[\begin{array} {ccccc}
a& b &\vline & e & f\cr
c& -a &\vline & g & -e\cr
\hline
0&0 &\vline &a& b\cr
0&0&\vline &c & -a
\end{array}\right].
\end{eqnarray}
 and solve the stationary zero curvature equation $\bar{W}_x=[\bar{U}, \bar{W}]$, namely,
\begin{eqnarray}\label{b2}
&& W_x=[U, W],\nonumber\\
&&  {W_1}_x=[U, W_1]+[U_1, W].
\end{eqnarray}
Obviously, the above equations become

 \begin{eqnarray}\label{b3}
&&a_x=pc-qb,\nonumber\\
&&  b_x=2(\lambda+h) b-2pa,\nonumber\\
&&   c_x=-2(\lambda+h) c+2qa,
\end{eqnarray}
as well as
 \begin{eqnarray}\label{b4}
&&  e_x=pg-qf+rc-sb,\nonumber\\
&& f_x=2(\lambda+h) f-2pe-2ra,\nonumber\\
&&   g_x=-2(\lambda+h) g+2 qe+2sa.
\end{eqnarray}
By assuming the following Laurent series expansions
\begin{eqnarray}\label{b5}
& & a=\sum_{i=0}^\infty a_i\lambda^{-i},\qquad b=\sum_{i=0}^\infty b_i\lambda^{-i},\qquad c=\sum_{i=0}^\infty c_i\lambda^{-i},\nonumber\\
& & e=\sum_{i=0}^\infty e_i\lambda^{-i},\qquad f=\sum_{i=0}^\infty f_i\lambda^{-i},\qquad g=\sum_{i=0}^\infty g_i\lambda^{-i},
\end{eqnarray}
and substituting \eqref{b5} into \eqref{b3} and \eqref{b4}, we arrive at
\begin{eqnarray}\label{b6}
& & {a_i}_x=pc_{i}-qb_{i},\nonumber\\
& & b_{i+1}=\frac{1}{2}{b_i}_x+pa_{i}-hb_i,\nonumber\\
& &  c_{i+1}=-\frac{1}{2}{c_i}_x+qa_{i}-hc_i, \qquad i\geq 0,
\end{eqnarray}
\begin{eqnarray}\label{b7}
& & {e_i}_x=pg_{i}-qf_{i}+rc_i-sb_i,\nonumber\\
& & f_{i+1}=\frac{1}{2}{f_i}_x+pe_{i}+ra_i-hf_i,\nonumber\\
& & g_{i+1}=-\frac{1}{2}{g_i}_x+qe_{i}+sa_i-hg_i, \qquad i\geq 0,
\end{eqnarray}
and
\begin{eqnarray}\label{b8}
& & b_0=c_0=f_0=g_0=0.
\end{eqnarray}
To guarantee the uniqueness of $\left\{a_i, b_i, c_i, e_i, f_i, g_i, ~i\geq 0\right\}$, we let $a_0=e_0=1$ and also need to impose the integration conditions
\begin{eqnarray*}
& & a_i|_{u=0}=b_i|_{u=0}=c_i|_{u=0}=0, \\
& & e_i|_{u=0}=f_i|_{u=0}=g_i|_{u=0}=0.
\end{eqnarray*}
Under the above assumptions, by means of the symbolic computation software Maple, we can obtain $\left\{a_i, b_i, c_i, e_i, f_i, g_i, ~i\geq 0\right\}$ explicitly.
The first four sets are listed as follows:
\begin{eqnarray*}
& & b_0=0, \quad  c_0=0,  \quad a_0=1,  \quad  f_0=0,   \quad  g_0=0,   \quad  e_0=1;\\
& & b_1=p, \quad  c_1=q,  \quad a_1=0,  \quad  f_1=p+r,   \quad  g_1=q+s,   \quad  e_1=0;\\
& & b_2=\frac{1}{2}p_x-\epsilon p(ps+qr), \quad  c_2=-\frac{1}{2}q_x-\epsilon q(ps+qr),   \quad a_2=-\frac{1}{2}pq,  \\
& & f_2=\frac{1}{2}p_x+\frac{1}{2}r_x-\epsilon (p+r)(ps+qr),   \quad g_2=-\frac{1}{2} q_x-\frac{1}{2} s_x-\epsilon (q+s)(ps+qr),\\
& & e_2=-\frac{1}{2}(pq+ps+qr);\\
& & b_3=\frac{1}{4}p_{xx}-\frac{1}{2}p^2q-\epsilon \left(\frac{1}{2}r_xpq+\frac{1}{2}s_xp^2+\frac{3}{2}p_xsp+rp_xq+\frac{1}{2}rpq_x\right)+\epsilon^2p(ps+qr)^2,\\
& & c_3=\frac{1}{4}q_{xx}-\frac{1}{2}pq^2+\epsilon \left(\frac{1}{2}r_xq^2+\frac{1}{2}s_xpq+\frac{1}{2}sp_xq+spq_x+\frac{3}{2}qrq_x\right)+\epsilon^2q(ps+qr)^2,\\
& & a_3=\frac{1}{4}pq_x-\frac{1}{4}p_xq+\epsilon pq (ps+qr),\\
& & f_3=\frac{1}{4}p_{xx}+\frac{1}{4}r_{xx}-\frac{1}{2}p^2q-\frac{1}{2}sp^2-rpq-\epsilon \left(\frac{1}{2}r_xpq+\frac{1}{2}rsp_x+\frac{1}{2}rps_x+\frac{3}{2}p_xsp\right.\\
& &  \qquad  \left.+\frac{3}{2}qrr_x+spr_x+\frac{1}{2}rpq_x+rp_xq+\frac{1}{2}r^2q_x+\frac{1}{2}s_xp^2\right)+\epsilon^2(p+r)(ps+qr)^2,\\
& & g_3=\frac{1}{4}q_{xx}+\frac{1}{4}s_{xx}-\frac{1}{2}pq^2-\frac{1}{2}rq^2-spq+\epsilon \left(\frac{1}{2}s_xpq+\frac{1}{2}rsq_x+\frac{1}{2}q^2r_x+\frac{3}{2}s_xsp\right.\\
& &  \qquad  \left.+\frac{1}{2}s^2p_x+\frac{1}{2}qsr_x+\frac{1}{2}sqp_x+spq_x+\frac{3}{2}qrq_x+qrs_x\right)+\epsilon^2(q+s)(ps+qr)^2,\\
& & e_3=\frac{1}{4}pq_x-\frac{1}{4}p_xq-\frac{1}{4}p_xs-\frac{1}{4}qr_x+\frac{1}{4}ps_x+\frac{1}{4}q_xr+\epsilon (pq+ps+qr)(ps+qr).
\end{eqnarray*}

The localness of the first four sets is not a coincidences. In fact, the functions $\left\{a_i, b_i, c_i, e_i, f_i, g_i, ~ i\geq 0\right\}$ are all local. 
First from $W_x=[U, W]$, we have
\begin{eqnarray*}
& & \frac{{\rm d}}{{\rm d}x}{\rm tr}\left(W^2\right)=2{\rm tr}(WW_x)=2{\rm tr}\left(W[U, W]\right)=0.
\end{eqnarray*}
Since ${\rm tr}(W^2)=2(a^2+bc)$, we can obtain
\begin{eqnarray*}
& & a^2+bc=\left(a^2+bc\right)\big|_{u=0}=1,
\end{eqnarray*}
based on the initial data \eqref{b8}. Then, by using the Laurent expansions \eqref{b5}, a balance of coefficients of $\lambda^i$ for each $i \geq 0$ tells that
\begin{eqnarray*}
& & a_{i+1}=-\frac{1}{2}\left(\sum_{j+k=i+1 \atop j,k\geq 1}a_ja_k+\sum_{j+k=i+1}b_jc_k\right).
\end{eqnarray*}
Similarly, we have
\begin{eqnarray*}
& & \frac{{\rm d}}{{\rm d}x}(2ae+fc+gb)=2a_xe+2ae_x+f_xc+fc_x+g_xb+gb_x\nonumber\\
& & \qquad \qquad\qquad\qquad =2(pc-qb)e+2a(pg+rc-sb-qf)\nonumber\\
& & \qquad \qquad\qquad\qquad\quad +[2(\lambda+h)f-2pe-2ra]c+f[-2(\lambda+h)c+2qa]\nonumber\\
& & \qquad \qquad\qquad\qquad\quad +[-2(\lambda+h)g+2qe+2sa]b+g[2(\lambda+h)b-2pa]\nonumber\\
& & \qquad \qquad\qquad\qquad =0.
\end{eqnarray*}
Thus we can obtain
\begin{eqnarray*}
& & 2ae+fc+gb=\left(2ae+fc+gb\right)\big|_{u=0}=2.
\end{eqnarray*}
Then, by means of the Laurent expansions \eqref{b5}, a balance of coefficients of $\lambda^i$ for each $i \geq 0$ tells that
\begin{eqnarray*}
& & e_{i+1}=-a_{i+1}-\sum_{j+k=i+1 \atop j,k\geq 1}a_je_k-\frac{1}{2}\sum_{j+k=i+1}f_jc_k-\frac{1}{2}\sum_{j+k=i+1}g_jb_k\nonumber\\
& & \qquad =\frac{1}{2}\sum_{j+k=i+1 \atop j,k\geq 1}a_ja_k+\frac{1}{2}\sum_{j+k=i+1}b_jc_k-\sum_{j+k=i+1 \atop j,k\geq 1}a_je_k-\frac{1}{2}\sum_{j+k=i+1}f_jc_k-\frac{1}{2}\sum_{j+k=i+1}g_jb_k.
\end{eqnarray*}
Based on the recursion relations \eqref{b6} and \eqref{b7}, an application of the mathematical induction finally shows that all functions $\left\{a_i, b_i, c_i, e_i, f_i, g_i, ~ i\geq 0\right\}$ are differential functions in $\bar{u}$, and so, they are all local.

Now, taking
\begin{eqnarray*}
& & \bar{V}^{[m]}=(\lambda^{m}\bar{W})_++\bar{\Delta}_m\\
& & \qquad=\left[\begin{array} {ccccc}
\sum_{i=0}^m a_i\lambda^{m-i}& \sum_{i=0}^m b_i\lambda^{m-i} &\vline & \sum_{i=0}^m e_i\lambda^{m-i} & \sum_{i=0}^m f_i\lambda^{m-i}\cr
\sum_{i=0}^m c_i\lambda^{m-i}& -\sum_{i=0}^m a_i\lambda^{m-i} &\vline & \sum_{i=0}^m g_i\lambda^{m-i} & -\sum_{i=0}^m e_i\lambda^{m-i}\cr
\hline
0&0 &\vline &\sum_{i=0}^m a_i\lambda^{m-i}& \sum_{i=0}^m b_i\lambda^{m-i}\cr
0&0&\vline &\sum_{i=0}^m c_i\lambda^{m-i} & -\sum_{i=0}^m a_i\lambda^{m-i}
\end{array}\right]\\
& & \quad\qquad+\left[\begin{array} {ccccc}
F_m&0 &\vline & 0 & 0\cr
0& -F_m &\vline & 0& 0\cr
\hline
0&0 &\vline &F_m& 0\cr
0&0&\vline &0& -F_m
\end{array}\right],
\end{eqnarray*}
the zero curvature equations
\begin{eqnarray*}
& & \bar{U}_{t_m}-\bar{V}_x^{[m]}+\left[\bar{U},\bar{V}^{[m]}\right]=0, \qquad m\geq 0
\end{eqnarray*}
give
\begin{eqnarray}
& & p_{t_m}=2b_{m+1}+2pF_m, \nonumber\\
& & q_{t_m}=-2c_{m+1}-2qF_m, \nonumber\\
& & r_{t_m}=2f_{m+1}+2rF_m, \nonumber\\
& & s_{t_m}=-2g_{m+1}-2sF_m, \nonumber\\
& &{F_m}_x=h_{t_m}.
\end{eqnarray}
Substituting the first four equations into the fifth one, we can compute
\begin{eqnarray*}
& &  {F_m}_x=h_{t_m}\\
& &  \qquad =\epsilon (p_{t_m}s+ps_{t_m}+q_{t_m}r+qr_{t_m})\\
& &  \qquad =\epsilon\left[(2b_{m+1}+2pF_m)s+p(-2g_{m+1}-2sF_m)+(-2c_{m+1}-2qF_m)r+q(2f_{m+1}+2rF_m)\right]\\
& &  \qquad =2\epsilon(sb_{m+1}-pg_{m+1}-rc_{m+1}+qf_{m+1})\\
& &  \qquad =-2 \epsilon {e_{m+1}}_x.
\end{eqnarray*}
Thus we introduce
\begin{eqnarray}
& & F_m=-2\epsilon e_{m+1},
\end{eqnarray}
and then  we have generated a complete system $\bar{u}_{t_m}=\bar{K}_m(\bar{u})$ of the AKNS integrable coupling:
\begin{eqnarray}\label{b11}
& &\left[\begin{array} {c}
p \cr
q \cr
r \cr
s
\end{array}\right]_{t_m}=\left[\begin{array} {c}
2b_{m+1}-4\epsilon pe_{m+1} \cr
-2c_{m+1}+4\epsilon qe_{m+1}\cr
2f_{m+1}-4\epsilon re_{m+1} \cr
-2g_{m+1}+4\epsilon se_{m+1}
\end{array}\right], \qquad m\geq 0.
\end{eqnarray}
  A nonlinear example in the above new system  is
\begin{eqnarray*}
& & p_{t_2}=-\frac{1}{2}p_{xx}-p^2q+\epsilon \left(pqp_x-2rp_xq-2p_xsp-2rpq_x-p^2q_x-2s_xp^2\right)\nonumber\\
& & \qquad\quad -2\epsilon^2 p(ps+qr)(qr+2pq+ps),\nonumber\\
& & q_{t_2}=-\frac{1}{2}q_{xx}+pq^2+\epsilon \left(qpq_x-q^2p_x-2sp_xq-2q^2r_x-2qrq_x-2spq_x\right)\nonumber\\
& & \qquad\quad +2\epsilon^2 q(ps+qr)(qr+2pq+ps),\nonumber\\
& & r_{t_2}=\frac{1}{2}r_{xx}+\frac{1}{2}p_{xx}-p^2q-p^2s-2rpq-\epsilon \left(rp_xq+3p_xsp+2r_xqr+r_xpq+2r_xps\right.\nonumber\\
& & \qquad \quad \left.+2r^2q_x+2rpq_x+2rps_x+s_xp^2\right)-2\epsilon^2 (ps+qr)(r^2q+pqr+rsp-sp^2),\nonumber\\
& & s_{t_2}=-\frac{1}{2}s_{xx}-\frac{1}{2}q_{xx}+pq^2+q^2r+2spq-\epsilon \left(spq_x+2sp_xq+2s^2p_x+q^2r_x+2qsr_x\right.\nonumber\\
& & \qquad \quad \left.+3qrq_x+2qrs_x+s_xpq+2sps_x\right)+2\epsilon^2 (ps+qr)(s^2p+pqs+rsq-rq^2).
\end{eqnarray*}
In the next section, we will show that this new generalized system \eqref{b11}  is bi-Hamiltonian.

\section{Bi-Hamiltonian structure}

 In this section, we will establish bi-Hamiltonian structures for the generalized \eqref{b11} by using the component-trace identity \eqref{a4}. It is direct to see
\begin{eqnarray*}
& & W\frac{\partial U_1}{\partial \lambda}+W_1\frac{\partial U}{\partial \lambda}=\left[\begin{array} {cc}
e  & -f \cr
g   & e
\end{array}\right],\qquad \qquad {\textrm{tr}}\left(W\frac{\partial U_1}{\partial \lambda}+W_1\frac{\partial U}{\partial \lambda}\right)=2e;\\
& & W\frac{\partial U_1}{\partial  p}+W_1\frac{\partial U}{\partial p}=\left[\begin{array} {cc}
\epsilon se  & 0 \cr
0   & g+\epsilon se
\end{array}\right],\qquad {\textrm{tr}}\left(W\frac{\partial U_1}{\partial p}+W_1\frac{\partial U}{\partial p}\right)=g+2\epsilon se;\\
& & W\frac{\partial U_1}{\partial  q}+W_1\frac{\partial U}{\partial q}=\left[\begin{array} {cc}
\epsilon re+f  & 0 \cr
0   & \epsilon re
\end{array}\right],\qquad {\textrm{tr}}\left(W\frac{\partial U_1}{\partial q}+W_1\frac{\partial U}{\partial q}\right)=f+2\epsilon re;\\
& & W\frac{\partial U_1}{\partial r}+W_1\frac{\partial U}{\partial r}=\left[\begin{array} {cc}
\epsilon qe  & 0 \cr
0   & c+\epsilon qe
\end{array}\right],\qquad {\textrm{tr}}\left(W\frac{\partial U_1}{\partial r}+W_1\frac{\partial U}{\partial r}\right)=c+2\epsilon qe;\\
& & W\frac{\partial U_1}{\partial  s}+W_1\frac{\partial U}{\partial s}=\left[\begin{array} {cc}
b+\epsilon pe  & 0 \cr
0   & \epsilon pe
\end{array}\right],\qquad {\textrm{tr}}\left(W\frac{\partial U_1}{\partial s}+W_1\frac{\partial U}{\partial s}\right)=b+2\epsilon pe.
\end{eqnarray*}
Now the corresponding  component-trace identity \eqref{a4}  becomes
\begin{eqnarray*}
& &  \frac{\delta}{\delta \bar{u}}\int 2e {\rm d}x=\lambda^{-\gamma}\frac{\partial}{\partial \lambda}\lambda^\gamma \left[\begin{array} {c}
g+2\epsilon se \cr
f+2\epsilon re \cr
c+2\epsilon qe \cr
b+2\epsilon pe
\end{array}\right].
\end{eqnarray*}
Balancing coefficients of each power of $\lambda$ in the above equality, we have
\begin{eqnarray*}
& &  \frac{\delta}{\delta \bar{u}}\int 2e_m {\rm d}x=(\gamma-m+1) \left[\begin{array} {c}
g_{m-1}+2\epsilon se_{m-1} \cr
f_{m-1}+2\epsilon re_{m-1} \cr
c_{m-1}+2\epsilon qe_{m-1} \cr
b_{m-1}+2\epsilon pe_{m-1}
\end{array}\right].
\end{eqnarray*}
Consider the particular case with $m=2$, we have $\gamma=0$. Therefore, we obtain
\begin{eqnarray}\label{c1}
& &  \frac{\delta}{\delta \bar{u}}\int \frac{-2e_{m+2}}{m+1} {\rm d}x= \left[\begin{array} {c}
g_{m+1}+2\epsilon se_{m+1} \cr
f_{m+1}+2\epsilon re_{m+1} \cr
c_{m+1}+2\epsilon qe_{m+1} \cr
b_{m+1}+2\epsilon pe_{m+1}
\end{array}\right].
\end{eqnarray}
In order to establish the relation between the new integrable hierarchy \eqref{b11} and the variational derivative formula \eqref{c1},
we first compute
\begin{eqnarray*}
& &  2b_{m+1}-4\epsilon pe_{m+1}=2(b_{m+1}+2\epsilon pe_{m+1})-8\epsilon pe_{m+1}\\
& &  \qquad\qquad\qquad\qquad =2(b_{m+1}+2\epsilon pe_{m+1})-8\epsilon p\partial^{-1}(pg_{m+1}+rc_{m+1}-sb_{m+1}-qf_{m+1})\\
& &  \qquad\qquad\qquad\qquad =2 (b_{m+1}+2\epsilon pe_{m+1})-8\epsilon p \partial^{-1}p(g_{m+1}+2\epsilon se_{m+1})-8\epsilon p \partial^{-1}r(c_{m+1}+2\epsilon qe_{m+1})\\
& &  \qquad\qquad\qquad\qquad\quad +8\epsilon p \partial^{-1}s(b_{m+1}+2\epsilon pe_{m+1})+8\epsilon p \partial^{-1}q(f_{m+1}+2\epsilon re_{m+1});
\end{eqnarray*}
\begin{eqnarray*}
& &  -2c_{m+1}+4\epsilon qe_{m+1}=-2(c_{m+1}+2\epsilon qe_{m+1})+8\epsilon qe_{m+1}\\
& &  \qquad\qquad\qquad\qquad =-2(c_{m+1}+2\epsilon qe_{m+1})+8\epsilon q\partial^{-1}(pg_{m+1}+rc_{m+1}-sb_{m+1}-qf_{m+1})\\
& &  \qquad\qquad\qquad\qquad =-2 (c_{m+1}+2\epsilon qe_{m+1})+8\epsilon q \partial^{-1}p(g_{m+1}+2\epsilon se_{m+1})+8\epsilon q \partial^{-1}r(c_{m+1}+2\epsilon qe_{m+1})\\
& &  \qquad\qquad\qquad\qquad\quad -8\epsilon q \partial^{-1}s(b_{m+1}+2\epsilon pe_{m+1})+8\epsilon q \partial^{-1}q(f_{m+1}+2\epsilon re_{m+1});
\end{eqnarray*}
\begin{eqnarray*}
& &  2f_{m+1}-4\epsilon re_{m+1}=2(f_{m+1}+2\epsilon re_{m+1})-8\epsilon re_{m+1}\\
& &  \qquad\qquad\qquad\qquad =2(f_{m+1}+2\epsilon re_{m+1})-8\epsilon r\partial^{-1}(pg_{m+1}+rc_{m+1}-sb_{m+1}-qf_{m+1})\\
& &  \qquad\qquad\qquad\qquad =2 (f_{m+1}+2\epsilon re_{m+1})-8\epsilon r \partial^{-1}p(g_{m+1}+2\epsilon se_{m+1})-8\epsilon r \partial^{-1}r(c_{m+1}+2\epsilon qe_{m+1})\\
& &  \qquad\qquad\qquad\qquad\quad +8\epsilon r \partial^{-1}s(b_{m+1}+2\epsilon pe_{m+1})+8\epsilon r \partial^{-1}q(f_{m+1}+2\epsilon re_{m+1});
\end{eqnarray*}
\begin{eqnarray*}
& &  -2g_{m+1}+4\epsilon se_{m+1}=-2(g_{m+1}+2\epsilon se_{m+1})+8\epsilon se_{m+1}\\
& &  \qquad\qquad\qquad\qquad =-2(g_{m+1}+2\epsilon se_{m+1})+8\epsilon s\partial^{-1}(pg_{m+1}+rc_{m+1}-sb_{m+1}-qf_{m+1})\\
& &  \qquad\qquad\qquad\qquad =-2(g_{m+1}+2\epsilon se_{m+1})+8\epsilon s \partial^{-1}p(g_{m+1}+2\epsilon se_{m+1})+8\epsilon s \partial^{-1}r(c_{m+1}+2\epsilon qe_{m+1})\\
& &  \qquad\qquad\qquad\qquad\quad -8\epsilon s \partial^{-1}s(b_{m+1}+2\epsilon pe_{m+1})-8\epsilon s \partial^{-1}q(f_{m+1}+2\epsilon re_{m+1}).
\end{eqnarray*}
Consequently, we obtain the following Hamiltonian structure for \eqref{b11}
\begin{eqnarray}\label{c2}
& & \bar{u}_{t_m}=\bar{K}_m=\bar{J}\frac{\delta \bar{\mathcal{H}}_m}{\delta \bar{u}},
\end{eqnarray}
with the Hamiltonian operator
\begin{eqnarray*}
& & \bar{J}=\left[\begin{array} {cccc}
-8\epsilon p\partial^{-1}p  & 8\epsilon p\partial^{-1}q & -8\epsilon p\partial^{-1}r & 2+8\epsilon p\partial^{-1}s\cr
8\epsilon q\partial^{-1}p  & -8\epsilon q\partial^{-1}q & -2+8\epsilon q\partial^{-1}r & -8\epsilon q\partial^{-1}s\cr
-8\epsilon r\partial^{-1}p & 2+8\epsilon r\partial^{-1}q & -8\epsilon r\partial^{-1}r &8\epsilon r\partial^{-1}s\cr
-2+8\epsilon s\partial^{-1}p & -8\epsilon s\partial^{-1}q &8\epsilon s\partial^{-1}r &-8\epsilon s\partial^{-1}s
\end{array}\right]
\end{eqnarray*}
and the Hamiltonian functionals
\begin{eqnarray*}
& &  \bar{\mathcal{H}}_m= \int \frac{-2e_{m+2}}{m+1} {\rm d}x, \qquad m\geq 0.
\end{eqnarray*}

It is now a direct computation to show that all members in the new integrable hierarchy \eqref{b11} are bi-Hamiltonian.
We compute the recursion operator $\Phi\equiv (\Phi_{ij})_{4\times 4}$ through
\begin{eqnarray*}
\left[\begin{array} {c}
2b_{m+1}-4\epsilon pe_{m+1} \cr
-2c_{m+1}+4\epsilon qe_{m+1}\cr
2f_{m+1}-4\epsilon re_{m+1} \cr
-2g_{m+1}+4\epsilon se_{m+1}
\end{array}\right]=\left[\begin{array} {cccc}
\Phi_{11}  & \Phi_{12} &\Phi_{13} &\Phi_{14}\cr
\Phi_{21}  & \Phi_{22} &\Phi_{23} &\Phi_{24}\cr
\Phi_{31}  & \Phi_{32} &\Phi_{33} &\Phi_{34}\cr
\Phi_{41}  & \Phi_{42} &\Phi_{43} &\Phi_{44}
\end{array}\right]\left[\begin{array} {c}
2b_{m}-4\epsilon pe_{m} \cr
-2c_{m}+4\epsilon qe_{m}\cr
2f_{m}-4\epsilon re_{m} \cr
-2g_{m}+4\epsilon se_{m}
\end{array}\right].
\end{eqnarray*}
Firstly, we have
\begin{eqnarray*}
& & 2b_{m+1}-4\epsilon pe_{m+1}\\
& & \quad={b_m}_x+2pa_m-2hb_m-4\epsilon p\partial^{-1}(pg_{m+1}+rc_{m+1}-sb_{m+1}-qf_{m+1})\\
& & \quad  ={b_m}_x+2pa_m-2hb_m-4\epsilon p \partial^{-1}p\left(\frac{1}{2}{g_m}_x+qe_m+sa_m-hg_m\right)-4\epsilon p\partial^{-1}r\left(-\frac{1}{2}{c_m}_x+qa_m-hc_m\right)\\
& & \qquad +4\epsilon p \partial^{-1}s\left(\frac{1}{2}{b_m}_x+pa_m-hb_m\right)+4\epsilon p\partial^{-1}q\left(\frac{1}{2}{f_m}_x+qe_m+ra_m-hf_m\right)\\
& & \quad  ={b_m}_x+2pa_m-2hb_m-4\epsilon p \partial^{-1}\left(-phg_m-rhc_m+shb_m+qhf_m\right)+2\epsilon p\partial^{-1}p{g_m}_x\\
& & \qquad +2\epsilon p\partial^{-1}r{c_m}_x+2\epsilon p\partial^{-1}s{b_m}_x+2\epsilon p\partial^{-1}q{f_m}_x\\
& & \quad  ={b_m}_x+2p\partial^{-1}(pc_m-qb_m)-2hb_m+4\epsilon p \partial^{-1}h{\partial} e_m+2\epsilon p\partial^{-1}p{g_m}_x+2\epsilon p\partial^{-1}r{c_m}_x\\
& & \qquad+2\epsilon p\partial^{-1}s{b_m}_x+2\epsilon p\partial^{-1}q{f_m}_x\\
& & \quad =\frac{1}{2}\partial (2b_m-4\epsilon pe_m)-p\partial^{-1} p (-2c_m+4\epsilon qe_m)-p\partial^{-1}q (2b_m-4\epsilon pe_m)-h(2b_m-4\epsilon pe_m)\\
& &\qquad -\epsilon p\partial^{-1} p \partial (-2g_m+4\epsilon se_m)-\epsilon p\partial^{-1} r \partial (-2c_m+4\epsilon qe_m)+\epsilon p\partial^{-1} s \partial (2b_m-4\epsilon pe_m)\\
& &\qquad +\epsilon p\partial^{-1} q\partial (2f_m-4\epsilon re_m)+2\epsilon \partial p \partial^{-1}(pg_m+rc_m-sb_m-qf_m)\\
& &\qquad +8\epsilon \partial p \partial^{-1}h(pg_m+rc_m-sb_m-qf_m)\\
& & \quad  =\left(\frac{1}{2}\partial-p\partial^{-1}q-h+\epsilon p \partial^{-1}s\partial-\epsilon \partial p \partial^{-1}s-4\epsilon p \partial^{-1} hs\right)(2b_m-4\epsilon p e_m)\\
& &\qquad +\left(-p\partial^{-1}p-\epsilon p\partial^{-1}r\partial-\epsilon\partial p\partial^{-1}r-4\epsilon p\partial^{-1}hr\right)(-2c_m+4\epsilon q e_m)\\
& &\qquad +\left(\epsilon p\partial^{-1}q\partial-\epsilon\partial p\partial^{-1}q-4\epsilon p\partial^{-1}hq\right)(2f_m-4\epsilon r e_m)\\
& &\qquad +\left(-\epsilon p\partial^{-1}p\partial-\epsilon\partial p\partial^{-1}p-4\epsilon p\partial^{-1}hp\right)(-2g_m+4\epsilon s e_m)\\
& & \quad  =\Phi_{11}(2b_m-4\epsilon p e_m)+\Phi_{12}(-2c_m+4\epsilon q e_m)+\Phi_{13}(2f_m-4\epsilon r e_m)+\Phi_{14}(-2g_m+4\epsilon s e_m),
\end{eqnarray*}
which tells
\begin{eqnarray*}
& & \Phi_{11}=\frac{1}{2}\partial-p\partial^{-1}q-h+\epsilon p \partial^{-1}s\partial-\epsilon \partial p \partial^{-1}s-4\epsilon p \partial^{-1} hs,
\\
&&
\Phi_{12}=
-p\partial^{-1}p-\epsilon p\partial^{-1}r\partial-\epsilon\partial p\partial^{-1}r-4\epsilon p\partial^{-1}hr,
\\
&&
\Phi_{13}=
\epsilon p\partial^{-1}q\partial-\epsilon\partial p\partial^{-1}q-4\epsilon p\partial^{-1}hq,
\\
&&
\Phi_{14}=
-\epsilon p\partial^{-1}p\partial-\epsilon\partial p\partial^{-1}p-4\epsilon p\partial^{-1}hp.
\end{eqnarray*}
Similarly, we have
\begin{eqnarray*}
& & \Phi_{21}=q\partial^{-1}q-\epsilon q\partial^{-1}s\partial+\epsilon\partial q\partial^{-1}s+4\epsilon q\partial^{-1}hs, \\
& & \Phi_{22}=-\frac{1}{2}\partial+q\partial^{-1}p-h+\epsilon q\partial^{-1}r\partial+\epsilon\partial q\partial^{-1}r+4\epsilon q\partial^{-1}hr, \\
& & \Phi_{23}=-\epsilon q\partial^{-1}q\partial+\epsilon\partial q\partial^{-1}q+4\epsilon q\partial^{-1}hq,\\
& & \Phi_{24}=\epsilon q\partial^{-1}p\partial+\epsilon\partial q\partial^{-1}p+4\epsilon q\partial^{-1}hp;\\
& & \Phi_{31}=-r\partial^{-1}q+\epsilon r\partial^{-1}s\partial-\epsilon\partial r\partial^{-1}s-p\partial^{-1}s-4\epsilon r\partial^{-1}hs,\\
& & \Phi_{32}=-r\partial^{-1}p-\epsilon r\partial^{-1}r\partial-\epsilon\partial r\partial^{-1}r-p\partial^{-1}r-4\epsilon r\partial^{-1}hr,\\
& & \Phi_{33}=\frac{1}{2}\partial-h+\epsilon r\partial^{-1}q\partial-\epsilon\partial r\partial^{-1}q-p\partial^{-1}q-4\epsilon r\partial^{-1}hq,\\
& & \Phi_{34}=-\epsilon r\partial^{-1}p\partial-\epsilon\partial r\partial^{-1}p-p\partial^{-1}p-4\epsilon r\partial^{-1}hp;\\
& & \Phi_{41}=s\partial^{-1}q-\epsilon s\partial^{-1}s\partial-\epsilon\partial s\partial^{-1}s+p\partial^{-1}s+4\epsilon s\partial^{-1}hs,\\
& & \Phi_{42}=s\partial^{-1}p+\epsilon s\partial^{-1}r\partial-\epsilon\partial s\partial^{-1}r+p\partial^{-1}r+4\epsilon s\partial^{-1}hr,\\
& & \Phi_{43}=-\epsilon s\partial^{-1} q\partial-\epsilon\partial s\partial^{-1}q+p\partial^{-1}q+4\epsilon s\partial^{-1}hq,\\
& & \Phi_{44}=-\frac{1}{2}\partial-h+\epsilon s\partial^{-1}p\partial-\epsilon\partial s\partial^{-1}p+p\partial^{-1}p+4\epsilon s\partial^{-1}hp.
\end{eqnarray*}
So we finally arrive at
\begin{eqnarray}\label{c3}
\bar{u}_{t_m}=\bar{K}_m=\bar{J}\frac{\delta \bar{\mathcal{H}}_{m}}{\delta \bar{u}}=\bar{M}\frac{\delta \bar{\mathcal{H}}_{m-1}}{\delta \bar{u}},\qquad m\geq 1,
\end{eqnarray}
where the second Hamiltonian operator $M$ is given by
\begin{equation}\label{c4}
\bar{M}=\Phi\bar{J}.
\end{equation}

So far, we are ready to see that
the new integrable hierarchy \eqref{b11} is integrable in the sense of Liouville.
That is, it possesses infinitely many independent commuting symmetries and conservation laws. In particular, we have the Abelian symmetry algebra of symmetries,
\begin{eqnarray*}
[\bar{K}_i,\bar{K}_j]=\bar{K}_i'(\bar{u})[\bar{K}_j]-\bar{K}_j'(\bar{u})[\bar{K}_i]=0,\qquad i,j\ge 0,
\end{eqnarray*}
and the Abelian algebras of conserved functionals,
\begin{eqnarray*}
\{\bar{\mathcal{H}}_i,\bar{\mathcal{H}}_j\}_{J}=\int \left( \frac{\delta\bar{\mathcal{H}}_i}{\delta \bar{u}}\right)^T \bar{J}\frac{\delta \bar{\mathcal{H}}_j}{\delta \bar{u}}{\rm d}x=0,\qquad i, j\ge 0,
\end{eqnarray*}
and
\begin{eqnarray*}
\{\bar{\mathcal{H}}_i,\bar{\mathcal{H}}_j\}_{M}=\int \left( \frac{\delta\bar{\mathcal{H}}_i}{\delta \bar{u}}\right)^T \bar{M}\frac{\delta \bar{\mathcal{H}}_j}{\delta \bar{u}}{\rm d}x=0, \qquad i, j\ge 0.
\end{eqnarray*}

\section{Conclusions and discussions}

 It is known that once a generating scheme associated with a non-semisimple Lie algebra is established, it can be used to
construct integrable couplings.
 The following non-semisimple Lie algebras formed by $2\times 2$,  $3\times 3$  and $4\times 4$  block matrices \cite{ma2011, ma-cam, mmz}
 \begin{eqnarray*}
& & \left [
\begin{array}{cccc}
  A_{1} & A_{2}\\
  0  &  A_{1}+A_2
\end{array}
\right],\qquad \left[\begin{array} {ccc}
A_1&A_2&A_3 \cr
0&A_1+\alpha A_2&\beta A_2+\alpha A_3 \cr
0&0&A_1+\alpha A_2
\end{array}\right],\\
& & \left[\begin{array} {cccc}
A_1&A_2&A_3&A_4 \cr
0&A_1+\alpha A_2&\alpha A_3 &\beta A_2+\alpha A_4\cr
0&0&A_1+\alpha A_2+\mu A_3 & \nu A_3 \cr
0&0&0&A_1+\alpha A_2
\end{array}\right]
\end{eqnarray*}
have been used to construct integrable couplings,  where $\alpha,\beta,\mu,\nu $ are arbitrary constants.
Certain kinds of  integrable couplings based on the above non-semisimple Lie algebras  have been obtained recently \cite{ma-maa,gz-jmp, zhang, gzy, ma-jmp,xyc,gz,mxz,mc,gz-pla, gz-csf2,st,gz-csf, mg-mplb, lf,ma-ra,zt,zht,zf,mz,mz-aa,ma-aip1,taz,ma-cam,  ma-aip2, sljy, yyzs}.
We have proposed the idea of using perturbation to construct new integrable systems, which generalizes the corresponding integrable couplings. As an example, the complete system of the AKNS integrable coupling, together with the recursion operator $\Phi$ and the bi-Hamiltonian structure \eqref{c3}, is generated successfully to illustrate the idea. The key step is that a perturbation term $h=\epsilon (ps+qr)$ is introduced and actually, the perturbation term could take a more generalized form $h=\sum_{j=1}^N\epsilon_j (ps+qr)_{jx}$. The resulting construction procedure can be applied to many other cases, including the Dirac, multi-component AKNS, WKI, KN, super-AKNS and Volterra spectral problems \cite{ma-maa,gz-jmp, zhang, gzy, ma-jmp,xyc,gz,mxz,mc,gz-pla, gz-csf2,st,gz-csf, mg-mplb, lf,ma-ra,zt,zht,zf,mz,mz-aa,ma-aip1,taz,ma-cam,  ma-aip2, sljy, yyzs, ma2011, mmz}.

In addition, we mention that finite-dimensional irreducible representations \cite{ma-aip1} of  some Lie algebras  can also be used  to create  integrable couplings.
For instance, a spectral matrix using $V_2$
\begin{eqnarray}
& & \phi_x=\bar{U}\phi,\qquad \bar{U}=\left[\begin{array} {ccccccc}
3\lambda & p &0&0&\vline & r &0\cr
3q & \lambda &2p&0&\vline & 0 &r\cr
0 & 2q &-\lambda &3p&\vline & s &0\cr
0 & 0 &q&-3\lambda&\vline & 0 &s\cr
\hline
0&0&0&0 &\vline & \lambda & p\cr
0&0&0&0&\vline &q & -\lambda
\end{array}\right]
\end{eqnarray}
could be another example.
Replacing $\lambda $ with $\lambda+h$ in the above matrix and setting
\begin{eqnarray}
& & \bar{W}=\left[\begin{array} {ccccccc}
3a & b &0&0&\vline & f &0\cr
3c & a &2b&0&\vline & e &f\cr
0 & 2c &-a &3b&\vline & g &e\cr
0 & 0 &c&-3a&\vline & 0 &g\cr
\hline
0&0&0&0 &\vline & a & b\cr
0&0&0&0&\vline &c & -a
\end{array}\right],
\end{eqnarray}
we can also construct new completion of the AKNS integrable coupling in the same manner. For convenience, we omit the construction process and the associated results.

\section*{Acknowledgements}

%We are appreciated to the referee and editor for their constructive suggestions and helpful comments.

This work is in part supported by the national natural science foundation of China (Grant No. 11371323, 11371326 and 11271266), Beijing Municipal Natural Science Foundation (Grant No. 1162003), NSF under the grant DMS-1664561, and the 111 project of China (B16002).

%---------------
\vspace{0.3cm}
%---------------


\begin{thebibliography}{sl}

\bibitem{mf-csf} W. X. Ma, B. Fuchssteiner, Integrable theory of the perturbation equations, Chaos, Solitons \& Fractals 7 (1996)
1227-1250.



\bibitem{ma-maa} W. X. Ma, Integrable couplings of soliton equations by perturbations I-A general theory and application to the
KdV hierarchy, Methods Appl. Anal. 7 (2000) 21-55.



 \bibitem{gz-jmp} F.K. Guo, Y.F. Zhang, A new loop algebra and a corresponding integrable hierarchy, as well as its integrable coupling, J. Math. Phys. 44 (2003) 5793-5803.

 \bibitem{zhang} Y.F. Zhang, A generalized multi-component Glachette-Johnson (GJ) hierarchy and its integrable coupling system, Chaos, Solitons and Fractals 21 (2004) 305-310.


\bibitem{gzy}  F.K. Guo, Y.F. Zhang, Q.Y. Yan, New simple method for obtaining integrable hierarchies of soliton equations with multicomponent potential functions,  Inter. J.  Theor. Phys. 43 (2004) 1139-1146.


\bibitem{ma-jmp} W.X. Ma,  Integrable couplings of vector AKNS soliton equations, J. Math. Phys. 46 (2005) 033507 (19pp).



\bibitem{xyc} T.C. Xia, F.J. Yu, D.Y. Chen, The multi-component generalized Wadati-Konono-Ichikawa (WKI) hierarchy and its
multi-component integrable couplings system with two arbitrary functions, Chaos, Solitons and Fractals 24 (2005) 877-883.


\bibitem{gz} F.K. Guo, Y.F. Zhang, The quadratic-form identity for constructing the
Hamiltonian structure of integrable systems, J. Phys. A: Math. Gen. 38 (2005) 8537-8548.



\bibitem{mxz} W.X. Ma, X.X. Xu, Y.F. Zhang,  Semidirect sums of Lie algebras and discrete integrable couplings, J. Math. Phys. 47 (2006) 053501 (16pp).




\bibitem{mc} W.X. Ma, M. Chen, Hamiltonian and quasi-Hamiltonian structures associated with semi-direct sums of Lie algebras, J. Phys. A: Math. Gen. 39 (2006) 10787-10801.



\bibitem{gz-pla} F.K. Guo, Y.F. Zhang, Two unified formulae, Phys. Lett. A 366 (2007) 403-410.



\bibitem{gz-csf2} F.K. Guo, Y.F. Zhang, The integrable coupling of the AKNS hierarchy and its Hamiltonian structure, Chaos, Solitons and Fractals 32 (2007) 1898-1902.



\bibitem{st} Y.P. Sun, Hon-Wah Tam, A hierarchy of non-isospectral multi-component AKNS equations and its integrable couplings, Phys. Lett. A 370 (2007) 139-144.

\bibitem{gz-csf} F.K. Guo, Y.F. Zhang, The computational formula on the constant $\gamma$ appeared in the
equivalently used trace identity and quadratic-form identity, Chaos, Solitons and Fractals 38 (2008) 499-505.


\bibitem{mg-mplb}  W.X. Ma, L. Gao, Coupling integrable couplings, Modern Phys. Lett. B  23  (2009) 1847-1860.



\bibitem{lf} L. Luo, E.G. Fan, The algebraic structure of discrete zero curvature
equations associated with integrable couplings and application to enlarged Volterra systems, Science in China Series A: Math. 52 (2009) 147-159.



\bibitem{ma-ra} W.X. Ma, Variational identities and applications to Hamiltonian structures of soliton equations, Nonlinear Analysis: Theory, Methods \& Applications 71 (2009) e1716-e1726.



\bibitem{zt} Y.F. Zhang, Hon-Wah Tam,  Three kinds of coupling integrable couplings of the Korteweg-de Vries hierarchy of
evolution equations, J. Math. Phys. 51 (2010) 043510 (18pp).



\bibitem{zht} Y.F. Zhang, Hon-Wah Tam,   Four Lie algebras associated with $R^6$ and their applications, J. Math. Phys. 51 (2010) 093514 (30pp).


\bibitem{zf}  Y.F. Zhang, E.G. Fan, Coupling integrable couplings and bi-Hamiltonian
structure associated with the Boiti-Pempinelli-Tu hierarchy, J. Math. Phys. 51 (2010) 083506 (18pp).



\bibitem{mz} W.X. Ma, Z.N. Zhu, Constructing nonlinear discrete integrable Hamiltonian couplings, Comput. Math. Appl. 60 (2010) 2601-2608.


\bibitem{mz-aa} W.X. Ma, Y. Zhang, Component-trace identities for Hamiltonian structures, Appl. Anal. 89 (2010) 457-472.


\bibitem{ma-aip1} W.X. Ma, Variational identities and Hamiltonian structures, in: W.X. Ma, X.B. Hu, Q.P. Liu (Eds.), Nonlinear and Modern Mathematical Physics, AIP Conference Proceedings, Vol. 1212 (American Institute of Physics, Melville, NY, 2010) 1-27.


\bibitem{taz} Hon-Wah Tam, Y.F. Zhang, An integrable system and associated integrable models as well as Hamiltonian
structures, J. Math. Phys. 53 (2012) 103508 (25pp).


\bibitem{ma-cam} W.X. Ma, Loop algebras and bi-integrable couplings, Chin. Ann. Math. Ser. B 33 (2012) 207-224.



\bibitem{ma-aip2} W.X. Ma, Integrable coupling and matrix loop algebras, in: W.X. Ma and D. Kaup (Eds.), Proceeding of the 2nd International Workshop on Nonlinear and Modern Mathematical Physics, AIP coference Proceedings, Vol. 1562 (American Institute of Physics, Melville, NY, 2013) 105-122.


\bibitem{sljy} S.F. Shen, C.X. Li, Y.Y. Jin, S.M. Yu, Multi-component integrable couplings for the Ablowitz-Kaup-Newell-Segur and
Volterra hierarchies, Math. Meth. Appl. Sci. 38 (2015) 4345-4356.


\bibitem{yyzs} S.M. Yu, Y.J. Ye, J. Zhang, J.Q. Song, Tri-integrable coupling of the Kaup Newell soliton hierarchy and Liouville integrability, Modern Phys. Lett. B 30 (2016) 1650277 (13pp).


\bibitem{akns} M.J. Ablowitz, D.J. Kaup, A.C. Newell, H. Segur, The inverse scattering transform-Fourier analysis for nonlinear problems, Stud. Appl. Math. 53 (1974) 249-315.


 \bibitem{tu-jmp}G.Z. Tu, The trace identity, a powerful tool for constructing the Hamiltonian structure of integrable systems, J. Math. Phys. 30 (1989) 330-338.

\bibitem{tu-jpa} G.Z. Tu, A trace identity and its applications to the theory of discrete integrable systems, J. Phys. A: Math. Gen. 23 (1990) 3903-3922.


\bibitem{tah} G.Z. Tu, R.I. Andrushkiw, X.C. Huang, A trace identity and its application to integrable systems of $1+2$ dimensions, J. Math. Phys. 32 (1991) 1900-1907.

\bibitem{ma2011} W.X. Ma, Nonlinear continuous integrable Hamiltonian couplings, Appl. Math. Compu. 217 (2011) 7238-7244.


\bibitem{mmz}
 W.X. Ma, J.H. Meng, H.Q. Zhang,
 Integrable couplings, variational identities and Hamiltonian formulations, Global J. Math. Sci. 1 (2012) 1-17.








\end{thebibliography}
\end{document}